\documentclass[]{pasj01}
\draft

\Received{}
\Accepted{}
 
\begin{document} 

\title{ 
Steady jet ejections from the innermost region of advection-dominated accretion flow around a black hole}

\author{Hajime \textsc{Inoue}\altaffilmark{1}}%
\altaffiltext{1}{Institute of Space and Astronautical Science, Japan Aerospace Exploration Agency, 3-1-1 Yoshinodai, Chuo-ku, Sagamihara, Kanagawa 252-5210, Japan}
\email{inoue-ha@msc.biglobe.ne.jp}


\KeyWords{accretion, accretion disks  --- stars: black holes --- stars: jets --- X-rays: binaries}

\maketitle

\begin{abstract}
We study ejection mechanisms for two kinds of steady jets: one observed from black hole binaries in the low/hard state and the other from SS433. 
The specific energy of the ejected gas is required to be positive for the jets to get to infinity, while that of the accreted gas is naively considered to be negative at the outermost boundary of the accretion flow.
To reconcile the opposite sign of the specific energies, we propose a situation where two layers exist in the accretion flow and one layer receives energy from the other sufficiently for the specific energy to be positive.
For the steady jets in the low/hard state, the accretion ring at the outermost end of the accretion flow is considered to yield two-layer flow in which a geometrically thick ADAF sandwiches a geometrically thin accretion disk and the thin disk is supposed to turn to another ADAF on the inner side.
The energy transfer is expected to occur through turbulent mixing between the two layers and the upper layer is discussed to have the positive specific energy large enough for the terminal velocity to be $\sim$ 0.1 $c$.
For the steady jets from SS433, a slim disk is argued to separate into two stratified layers due to the photon diffusion in the direction perpendicular to the equatorial plane under the advection dominated situation.
In this case, the specific energy of the upper layer is expected to be positive such that the terminal velocity exceeds 0.2 $c$.
The jet ejection process near the black hole is investigated commonly to both the two-layer cases and predicts the jet opening angle becomes as small as 2$^{\circ}$.

\end{abstract}

\section{Introduction}\label{Introduction}
Relativistic jets are often observed from binary X-ray sources and active galactic nuclei (AGN).
The jet features were first found from the nuclei of galaxies and such observational evidences as superluminal motions indicate that the jet velocity is close to the velocity of light, $c$ (see e.g. Bridle \& Perley 1984 for the review on extragalactic radio jets).
Relativistic jets from Galactic binary sources was first discovered from SS433 by detecting Doppler drifts in the optical lines, which is naturally explained by  precessing motions of jets collimated to two opposite directions with velocity of 0.26 $c$ (see Margon 1984 for the early review).
In early 1990s, multi-wavelength studies of X-ray and gamma-ray sources in the Galactic center region discovered that some X-ray persistent sources exhibited radio images consisting of compact components at the center and two-sided jets and, furthermore, superluminal motions were  detected from some recurrent transient X-ray sources (see Mirabel \& Rodriguez 1999 for the review of relativistic jet sources in the Galaxy).  

These discoveries stimulated observational studies of relativistic jets in transient X-ray sources, and, based on them,
Fender, Belloni and Gallo (2004) presented a unified model for the disk-jet coupling in black hole X-ray binaries, suggesting existence of two types of jets: steady jets and transient jets.

The transient jets are always observed around the peak of an X-ray outburst and superluminal motions are often detected.
The sources around the outburst peak are considered to be in the very high state which is characterized by the various time variabilities (see e.g. section 5 of Inoue 2022a for the review of X-ray observations of accretion disks and references therein).
The various time variabilities are understood to be basically governed by the limit cycle behavior between the gas-pressure dominated standard disk and the slim disk.
The transient jets are discussed to appear at the moment when the innermost front of the slim disk extending from the outside reaches the boundary with the black hole in the cyclic behavior (see subsection 6.1 of Inoue 2022a).

The steady jets persist in the low/hard state, characterized by the radio emissions which probably arise in the synchrotron emission from conical, partially absorbed jets (Fender 2001).
A positive correlation between the radio flux and the X-ray flux is found for several sources (e.g. Corbel et al. 2003; Gallo et al. 2003).
The Lorentz factor of the jet velocity is estimated to be less than two and is significantly smaller that those of the transient jets (Fender et al. 2004).

The steady jets are also observed from SS433, but the source is considered to be not in the low/hard state but in a special state under the supercritical accretion rate (e.g. Margon 1984; Fabrika 2004; see also Inoue 2022b).
The steady jets from SS433 should be regarded as another type of steady jets than those in the low/hard state.

These observations reveal that the relativistic jets appear in the particular states of accretion flow, which strongly indicates that the jets arise in association with mass accretion on to a compact object at the activity center.
Although it is unclear yet whether SS433 is a neutron star source or a black hole source, we will hereafter advance considerations on the two types of the steady relativistic jets: the steady jets in the low/hard state, and the steady jets under the supercritical accretion, on the assumption that the central object is a black hole.
The transient jets will be discussed in a separate paper.

Recently, Inoue (2021a) discussed that matter flowing into the gravitational field of a black hole from the companion star in a close binary once forms a geometrically thick accretion ring around the black hole, and then the ring extends two layer accretion flow in which an optically thick and geometrically thin accretion disk (the standard disk: Shakura \& Sunyaev 1973) is sandwiched by an optically thin and geometrically thick advection-dominated accretion flow (ADAF: Narayan \& Yi 1994).
Inoue (2022a) further argued that the observations need three steady states for configurations of the accretion flow from the outermost accretion ring to the innermost stable circular orbit around a black hole.

The basic configuration of the accretion flow could be a situation that the two parallel flows from the outermost part of the accretion disk extends to the innermost stable circular orbit, called as the standard disk configuration.
The high/soft state can be considered to be in this configuration.

The configuration on the lower accretion rate side of the standard disk configuration is such that an optically gray ADAF appears on the down stream side of the truncated standard disk sandwiched by the optically thin ADAF, called as the ADAF configuration.
Several observational evidences indicate that the low/hard state is in this configuration.

The configuration on the higher accretion rate side of the standard disk configuration is such that an optically thick ADAF (the slim disk: Abramovicz et al. 1988) exists on the down stream side of the truncated gas-pressure-dominated standard disk sandwiched by the optically thin ADAF, called as the slim disk configuration.
This configuration happens in the very high state.

The terminal velocities of the jets are close to the velocity of the light, which suggests that the jets come from the vicinity of the event horizon of the black hole. 
Then, the steady jets from SS433 are likely to appear in the boundary region of the slim disk to the black hole, while the steady jets in the low/hard state could arise in the innermost region of the optically gray ADAF. 
Since the slim disk is a kind of ADAF, we can say that the boundary region of ADAF to the black hole would play a key role on ejections of the relativistic jets.

The equation for specific energy of matter in the ADAF is generally given as
\begin{equation}
\frac{1}{2} v^{2} + \frac{1}{2} \frac{\ell^{2}}{r^{2}} + \frac{1}{\gamma-1} c_{\rm s}^{2} -\frac{GM}{r-R_{\rm s}} -(\ell - \ell_{\rm in})\Omega = e,
\label{eqn:EnEq-ADAF}
\end{equation}
by approximately averaging the quantities over the $z$ direction at $r$.
Here and hereafter, we introduce the cylindrical coordinate ($r$, $\phi$, $z$), where the $z$ direction is the angular momentum axis of the accretion flow, and the radial distance $R$ as $R=\sqrt{r^{2} + z^{2}}$.
$v$ is the inflowing velocity, and $c_{\rm s}$ is the sound velocity defined as $c_{\rm s}^{2} = \gamma P/\rho$.  Here, we adopt the polytropic relation between the pressure, $P$, and the density, $\rho$, as 
\begin{equation}
P = K \rho^{\gamma}, 
\label{eqn:P-rho}
\end{equation}
where $K$ is the proportional constant.
We also adopt the pseudo-Newtonian potential (Paczynsky \& Wiita 1980) as
\begin{equation}
\Phi(R) = -\frac{GM}{R-R_{\rm s}},
\label{eqn:PseudoPotential}
\end{equation}
where $G$ and $M$ are the gravitational constant and the black hole mass, respectively, and $R_{\rm s}$ is the Schwarzschild radius.
$\ell$ and $\Omega$ are the specific angular momentum and the angular velocity of the inflowing matter, respectively.  $\ell_{\rm in}$ is the value of $\ell$ at the innermost radius of the ADAF.
$e$ is the specific total energy of the infalling matter given at the outer boundary.

Our concern in this paper is on jet ejections from the innermost regions of the ADAFs in the slim disk configuration and in the ADAF configuration and those ADAFs are considered to evolve from the thin standard disk in the outer regions.
Hence, $e$ is expected to be negative in these cases.
Furthermore, 
according to the studies of the accretion ring by Inoue (2021a; 2021b), the specific total energy, $e$, is approximately estimated to be $-GM/r_{\rm ar}$ at the distance of the accretion ring, $r_{\rm ar}$, even for the geometrically thick accretion flow (see appendix in Inoue 2021b).
Hence, we adopt the following premise to the present study as
\begin{equation}
e < 0.
\label{eqn:e_ob<0}
\end{equation}

The specific total energy $e$ consists of two components, as seen from equation (\ref{eqn:EnEq-ADAF}), 
the matter transport energy $e_{\rm b}$ (often called the Bernoulli parameter) 
and the viscous transport energy $e_{\rm v}$ respectively defined as
\begin{equation}
e_{\rm b} = \frac{1}{2} v^{2} + \frac{1}{2} \frac{\ell^{2}}{r^{2}} + \frac{1}{\gamma-1} c_{\rm s}^{2} -\frac{GM}{r-R_{\rm s}},
\label{eqn:e_m}
\end{equation}
and 
\begin{equation}
e_{\rm v} = (\ell - \ell_{\rm in})\Omega.
\label{eqn:e_v}
\end{equation}
Since the following relation holds from equation (\ref{eqn:EnEq-ADAF}) as
\begin{equation}
e_{\rm b} = e + e_{\rm v},
\label{eqn:e_b-e_o+e_v}
\end{equation}
the Bernoulli parameter $e_{\rm b}$ can be positive in the typical region of ADAF with $e_{\rm v} > 0$ even if $e < 0$, which was pointed out by Narayan and Yi (1994).

The positive Bernoulli parameter means that the accreted matter is possible to escape from the gravitational potential.
Utilizing the possibility, Blandford and Begelman (1999) proposed a model, called as advection-dominated inflow-outflow solutions (ADIOS), in which a small fraction of the supplied matter actually falls on to the black hole and the remain is driven away in the form of a wind by the viscous transport energy.
Although further studies have been done in the extension of this idea (e.g. Blandford \& Begelman 2004; Ferreria et al. 2006), a physical mechanism for the positive energy to change the inflow to the outflow is yet unclear and 
no obvious reason seems to exist why the place of the mass ejection is biased to the vicinity of the black hole, making the relativistic jets.

In order to understand the ejection mechanism of the relativistic jets, it should be necessary to know properties of the innermost region of the accretion flow on to the black hole.
The flow velocity should be the velocity of the light, $c$, at the event horizon, while the sound velocity is limited to be $c/\sqrt{3}$ at largest.
Hence, the innermost accretion flow should be supersonic, and pass through the transonic point somewhere before reaching the event horizon, since the accretion flow on the upstream side where the angular momentum is transfered outward should be subsonic.
Behaviors of the transonic flow have been studied in many papers too (see e.g. chapter 8 of Kato et al. 2008; a brief review in the introduction section of Kumar \& Chattopadhyay 2013).
An interesting argument among them is that a shock appears between two critical points from a subsonic flow to a supersonic flow and a mass outflow is induced behind the shock (see Kumar \& Chattopadhyay 2013; 2017 and references therein). 
The simulated solutions for mass ejection behind the shock are, however, obtained only when $e$ is slightly positive (e.g. Kumar \& Chattopadhyav 2013), and are not simply applicable to the present study under the premise in equation (\ref{eqn:e_ob<0}).

In this paper, we study two cases of the two-layer situations of the accretion flow in which a part of specific energy of one layer is transfered to the other and 
the specific energy of the energy-transfered laye becomes positive so as for mass ejection to be possible.

\begin{figure}
\begin{center}
  \includegraphics[width=8cm]{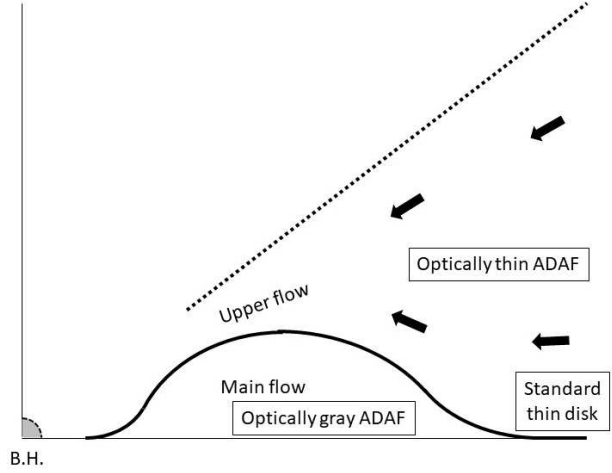} 
\end{center}
\caption{Schematic diagram of the ADAF configuration.  The standard thin disk is sandwiched by the optically thin ADAF on the outer side, and the standard thin disk changes to the optically gray ADAF on the inner side, where the optically thin ADAF runs on the optically gray ADAF.  We call the optically gray ADAF and the optically thin ADAF on the inner side as the main flow and the upper flow respectively.}
\label{fig:two-layer-configuration}
\end{figure}

One case is a two-layer situation in which an optically thin ADAF flowing from far outside runs on the upper side of an optically gray ADAF or a slim disk in the ADAF configuration or the slim disk configuration respectively.
Hereafter we call the flows on the equatorial plain side and the upper side as the main flow and the upper flow respectively.
Figure \ref{fig:two-layer-configuration} exhibits the schematic diagram of the two-layer situation expected in the ADAF configuration.
Such a configuration was already drawn in figure 9-(c) in Kumar and Yuan (2021)
as the result of the theoretical works.

The other case of the two-layer situation is discussed to appear in the main flow in the slim-disk configuration as the result of the photon diffusion from the bottom layer to the upper layer.

Then, we investigate how jets are ejected by utilizing the positive specific energy of the upper flow, commonly to the two cases of the two-layer situation.


\section{Transonic conditions from the outer subsonic flow to the inner supersonic flow}\label{InnermostBoundary}
We first review transonic conditions from the outer subsonic flow to the inner supersonic flow in an ADAF, to give bases for the study of steady jet ejections done in this paper.
The results in this section are consistent with the relevant works in the literature.

Assumptions of the study are as follows: 
First, the total specific energy of the flow, $e$, is conserved over the entire flow, assuming that the radiative energy loss is negligibly small.
Then, equation (\ref{eqn:e_ob<0}) is set as the premise on $e$ as discussed earlier.
Next, it is further premised that the inner supersonic flow is inviscid and the specific angular momentum is kept constant through the flow.
In the current study of the ADAF, the equation of angular momentum transfer is diffusive in form and signals can propagates to the up-stream side even in the supersonic flow (see e.g. Narayan et al. 1997).
However, this treatment in the case of the turbulence-dominated viscosity is problematic from the viewpoint of the causality (Kato 1994 and references therein).
Hence, we simply assume that the inner supersonic flow is inviscid.

We approximate that the supersonic flow is in a layer on a hollow cone surface with a constant $\theta$, where $\theta$ is an elevation angle from the equatorial plane of a position as viewed from the center.

If the flow becomes supersonic at the distance, $R_{\rm in}$, of the innermost boundary of the outer subsonic flow, the specific angular momentum can be considered to be conserved at $\ell_{\rm in}$ on the inner side than $R_{\rm in}$.
Thus, the dynamical equation of the inner supersonic flow along a hollow cone layer is expressed as
\begin{equation}
v\frac{dv}{dR} + \frac{1}{\rho}\frac{dP}{dR}+\frac{GM}{(R - R_{\rm s})^{2}} - \frac{\ell_{\rm in}^{2}}{r^{3}} \cos \theta = 0,
\label{eqn:DynamicalEq}
\end{equation}
and the energy equation is given as
\begin{equation}
\frac{1}{2} v^{2} + \frac{1}{2} \frac{\ell_{\rm in}^{2}}{r^{2}}+\frac{1}{\gamma-1} c_{\rm s}^{2} - \frac{GM}{R - R_{s}}  = e_{\rm in}.
\label{eqn:InnerBernoulliEq}
\end{equation}
Here, the parameter $e_{\rm in}$ is the specific total energy of the inflowing matter at $R_{\rm in}$ and is conserved through the supersonic flow inner than $R_{\rm in}$.
Since the specific angular momentum in the inner supersonic flow than $R_{\rm in}$ is conserved at $\ell_{\rm in}$, the viscous transport energy $e_{\rm v}$ should be zero at $R_{\rm in}$ as seen from equation (\ref{eqn:e_v}).
Then, considering a possibility for an energy increment transported from the adjacent layer, $e_{\rm in}$ is expressed as
\begin{equation}
e_{\rm in} = e + \Delta e,
\label{eqn:e_in=e_ob+De}
\end{equation}
where $\Delta e$ is the energy increment if it exists.

The continuity equation is written as
\begin{equation}
\rho v S = A,
\label{eqn:ContinuityEq}
\end{equation}
where $S$ is the cross section of the layer and $A$ is the constant representing the mass flow rate through it.
Differentiating  equation (\ref{eqn:ContinuityEq}) with $R$ on the assumption of $S \propto R^{2}$, and combining the result and equation (\ref{eqn:DynamicalEq}) with the help of the polytropic relation, we have the following equation for the transonic condition as
\begin{equation}
(v^{2} - c_{\rm s}^{2})\frac{1}{v}\frac{dv}{dR} = \frac{2 c_{\rm s}^{2}}{R} -\frac{GM}{(R - R_{\rm s})^{2}} + \frac{\ell_{\rm in}^{2}}{r^{3}} \cos \theta.
\label{eqn:TransonicCond}
\end{equation}
Since the right side of the above equation should be zero at the sonic point, we have the equation to determine the positions of the sonic point as
\begin{equation}
\frac{2 c_{\rm s}^{2}}{R} - \frac{GM}{(R - R_{\rm s})^{2}} + \frac{\ell_{\rm in}^{2}}{r^{3}} \cos \theta = 0.
\label{eqn:TrnSncCond_Rin_0}
\end{equation}
The sound velocity is calculated from equation (\ref{eqn:InnerBernoulliEq}) by equating $v = c_{\rm s}$ as
\begin{equation}
c_{\rm s} = \Gamma \left(e_{\rm in} + \frac{GM}{R - R_{\rm s}} - \frac{1}{2} \frac{\ell_{\rm in}^{2}}{r^{2}}\right),
\label{eqn:c_s_inner}
\end{equation}
where
\begin{equation}
\Gamma =\left(\frac{1}{2} + \frac{1}{\gamma-1}\right)^{-1} = \frac{2(\gamma-1)}{\gamma+1},
\label{eqn:Gamma_1}
\end{equation}
and equation (\ref{eqn:TrnSncCond_Rin_0}) is rewritten as
\begin{equation}
 2\Gamma \frac{e_{\rm in}}{R} - \frac{GM}{(R - R_{\rm s})^{2}R} \{R - 2\Gamma (R - R_{\rm s})\} + (1 - \Gamma) \frac{\ell_{\rm in}^{2}}{r^{3}}  \cos \theta = 0.
\label{eqn:TrnSncCond_Rin_1}
\end{equation}

\begin{figure}
\begin{center}
  \includegraphics[width=8cm]{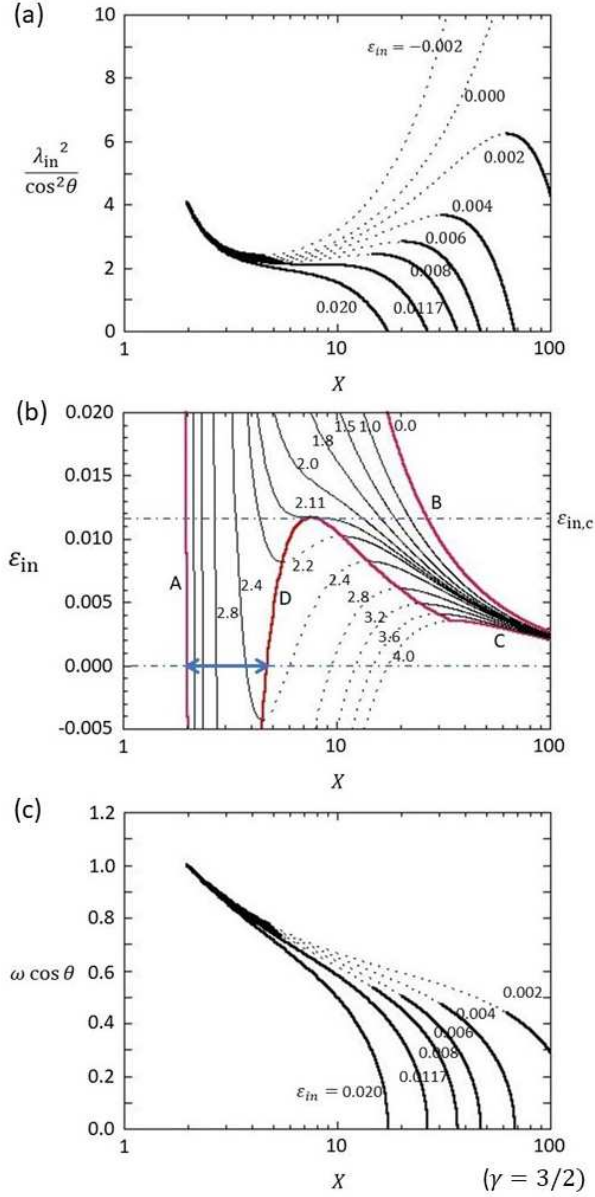} 
\end{center}
\caption{Diagrams showing normalized positions $X$ of the sonic points and their properties for inviscid and adiabatic flows, having the normalized specific energy $\varepsilon_{\rm in}$ and the normalized specific angular momentum $\lambda_{\rm in}$, along hollow cone surfaces with the elevation angle $\theta$, in case of $\gamma = 3/2$.  (a) Relations between  $\lambda_{\rm in}^{2}/\cos^{2} \theta$ and $X$ for given $\varepsilon_{\rm in}$; (b) Relations between $\varepsilon_{\rm in}$ and $X$ for given $\lambda_{\rm in}^{2}/\cos^{2} \theta$; and (c) Diagram indicating the reduction factor $\omega$ of the specific angular momentum from that of the Keplerian circular motion at the position of ($X$, $\theta$).  The portions of the thin solid line and the dotted line indicate the type of the sonic points as the saddle type and the center type respectively.  The thick solid lines in (b) represent the boundaries of the region in which the saddle-type sonic point exists.  The boundaries, A to D are characterized as follows, A: the constraint from equation (\ref{eqn:X_imp}); B: the case of the Bondi accretion ($\lambda_{\rm in}=0$), C: the constraint from equation (\ref{eqn:varepsilon_imp}) ;D: the boundary between the saddle-type region and the center type region.  The arrow in (b) exhibits the range of $X$ for the transonic point of the main flow, which is discussed in subsubsection \ref{MainFlow}.}
\label{fig:g=3per2}
\end{figure}

From this equation, we obtain the following equation as
\begin{equation}
\frac{\lambda_{\rm in}^{2}}{\cos^{2} \theta} = \frac{1}{2(1-\Gamma)} \frac{X^{2}}{(X-1)^{2}} \{ -2 \Gamma \varepsilon_{\rm in} (X-1)^{2} + (1-2\Gamma) X +2 \Gamma \},
\label{eqn:Lambda-Eq}
\end{equation}
where we have introduced non-dimensional parameters $X$ for the position of the sonic point, $\lambda$ for the specific angular momentum and $\varepsilon$ for the specific energy, defined as $X = R/R_{\rm s}$, $\lambda = \ell/(R_{\rm s}c)$ and $\varepsilon = e/(GM/R_{\rm s})$ respectively.
The values of $\lambda_{\rm in}^{2}/\cos^{2} \theta$ calculated from this equation are plotted against $X$ in figure \ref{fig:g=3per2}-(a) for various $\varepsilon_{\rm in}$ values in case of  $\gamma$ = 3/2.

Here, the innermost point of ($X_{\rm imp}$, $\lambda_{\rm imp}$) is obtained from the necessary condition that the sum of the kinetic energy and the thermal enthalpy cannot be negative in equation (\ref{eqn:InnerBernoulliEq}), which is written as
\begin{equation}
\frac{\lambda_{\rm in}^{2}}{\cos^{2} \theta} \leq \frac{X^{2}}{X-1} \{ \varepsilon_{\rm in} (X-1)+1\}.
\label{eqn:Cond-lambda}
\end{equation}
From the above two equations, the values of $X_{\rm imp}$ and $\lambda_{\rm imp}$ are calculated as
\begin{equation}
X_{\rm imp} \simeq 2 ( 1 - \varepsilon_{\rm in}),
\label{eqn:X_imp}
\end{equation}
and 
\begin{equation}
\frac{\lambda_{\rm imp}^{2}}{\cos^{2} \theta} \simeq 4 ( 1 + \varepsilon_{\rm in}),
\label{eqn:varepsilon_imp}
\end{equation}
when $|\varepsilon_{\rm in}| \ll 1$.

This figure tells us the positions of the sonic points and also the types of them.
This trend as shown in this figure was already studied by several authors (see chapter 8 of Kato et al. 2008 and references therein) and it is shown that the portions of the solid lines and the dashed lines in figure \ref{fig:g=3per2} correspond to the saddle-type sonic points and the the center-type sonic points respectively.
The flow can smoothly pass through the saddle-type sonic point but not through the center-type.
The center-type sonic point appears due to the effect of the centrifugal force on the flow.
On the outer side than the center type sonic point, the effect of the gravitational force is dominant to that of the centrifugal force, but the latter becomes dominant on the inner side.
As the result, the situation to satisfy equation (\ref{eqn:TrnSncCond_Rin_1}) is mathematically realized.
This center type sonic point cannot, however, be the transonic point between the subsonic region and the supersonic region of the flow.

In parallel to the $X$ - $\lambda_{\rm in}^{2}/\cos^{2} \theta$ plots, equation (\ref{eqn:TrnSncCond_Rin_1}) also gives us plots of $\varepsilon_{\rm in}$ against $X$ for given $\lambda_{\rm in}^{2}/\cos^{2} \theta$ values, which are shown in figure \ref{fig:g=3per2}-(b).
We can clearly see the region where the saddle-type sonic point exists on the $X$ - $\varepsilon_{\rm in}$ plane.

In case of $\varepsilon_{\rm in} \leq 0$, the saddle-type sonic point appears only in the region of $X$ from 2 to $\sim$ 5.
Hence, the viscid ADAF from the far outside in which the angular momentum is transfered outward should change to the inviscid supersonic flow through the inner saddle-type sonic point very close to the black hole.

The case of $\varepsilon_{\rm in} > 0$ is further divided into two sub-cases.

When $\varepsilon_{\rm in}$ is larger than the critical value, $\varepsilon_{\rm in,\;  c}$, which is 0.012 for $\gamma = 3/2$, only one saddle-type sonic point appears in the fairly large range of $X$.
The position of the transonic point goes inward as the specific angular momentum increases from $\lambda_{\rm in}^{2} = 0$ at the outermost point to $\lambda_{\rm in}^{2}/\cos^{2}/\cos^{2} \theta = 4$ at the innermost point.
The outer and inner side of the range can be called as the Bondy type and the disk type respectively (see section 8.1 in Kato et al. 2008).

\begin{figure}
\begin{center}
  \includegraphics[width=8cm]{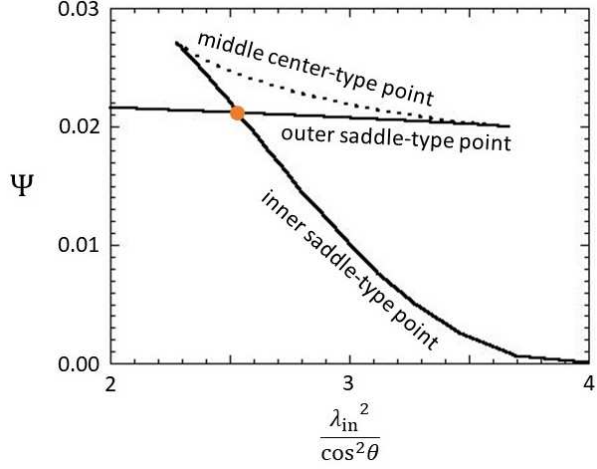} 
\end{center}
\caption{Loci of the sonic point solutions given by equation (\ref{eqn:Lambda-Eq}) on the $\lambda_{\rm in}^{2}/\cos^{2} \theta$ - $\Psi$ plane, in case of $\varepsilon_{\rm in}$ = 0.004 and $\gamma$ = 3/2.  One cross point appears between the loci of the inner saddle-type points and the outer saddle-type points}
\label{fig:L2-Psi}
\end{figure}

When $\varepsilon_{\rm in,\; c} \geq \varepsilon_{\rm in} > 0$, the region where the saddle-type sonic point exists are separated into two regions, the inner region and the outer region.
Although the two regions simultaneously appear for the same $\varepsilon_{\rm in}$, it is because equation (\ref{eqn:TrnSncCond_Rin_1}) is obtained by differentiating the continuity equation (\ref{eqn:ContinuityEq}) and does not mean that the inner and outer sonic points simultaneously appear in every practical flow.
Considering $c_{\rm s}^{2} = \gamma K \rho^{\gamma - 1}$ and $v = c_{\rm s}$ at the sonic point, we can have the following two non-dimensional equations from equations  (\ref{eqn:ContinuityEq}) and (\ref{eqn:InnerBernoulliEq}), which should be satisfied at the sonic points given by equation (\ref{eqn:Lambda-Eq}), as
\begin{equation}
\Psi = \Xi^{1/\Gamma} X^{2},
\label{eqn:Psi}
\end{equation}
and
\begin{equation}
\Xi= \Gamma \left( \varepsilon_{\rm in} + \frac{1}{X-1} - \frac{\lambda_{\rm in}^{2}/\cos^{2} \theta}{X^{2}} \right),
\label{eqn:beta}
\end{equation}
where $\Psi$ is the parameter of a combination of $A$, $K$ and $\gamma$ and $\Xi$ is defined to be $v^{2}/(GM/R_{\rm s})$.
We can calculate the $\Psi$ value from the above two equations for given values of $\lambda_{\rm in}^{2}/\cos^{2} \theta$, $\varepsilon_{\rm in}$ and $X$, and have a locus on the $\lambda_{\rm in}^{2} / \cos^{2} \theta$ and $\Psi$ plane for a given $\varepsilon_{\rm in}$ by changing the $X$ value.
A typical example is shown in figure \ref{fig:L2-Psi} and we see that the inner and outer saddle-type sonic points simultaneously appear only at the cross point of the respective loci for the inner and outer transonic points and the other flows have either of the two.
Fukue (1987) shows that when the specific angular momentum is larger than the value at the cross point, the supersonic flow to the black hole is realized through the inner sonic point, and that, on the other hand,  when the specific angular momentum is smaller than the cross point value, the supersonic flow is realized through the outer sonic point.

\begin{figure}
\begin{center}
  \includegraphics[width=8cm]{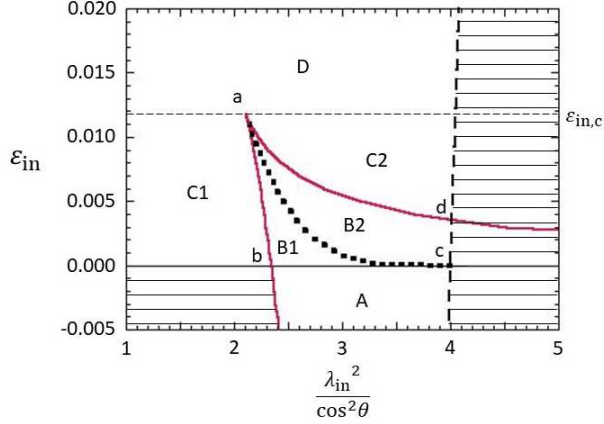} 
\end{center}
\caption{Domains for the characteristics of the sonic-point solutions on the $\lambda_{\rm in}^{2}/\cos^{2} \theta$ - $\varepsilon_{\rm in}$ plane.  The hatched region on the right hand side is the forbidden region given by the discussion on equation (\ref{eqn:varepsilon_imp}).  The other hatched region on the left, bottom corner is the no-solution region.  This region in which the specific angular momentum is very low is identical to the case of the Bondi accretion in which no solution exists when the specific energy is negative. As the result, the transonic flow is realized only in the region A in which the specific angular momentum is high enough for the sonic point to appear even when $\varepsilon_{\rm in}< 0$. Three (inner, middle and outer) sonic-point solutions are obtained in the region surrounded by the a-b-c-d-a lines, in which the transonic flow through the outer point is realized in the region B1 and that through the inner point happens in the region B2.  Including the regions of C1 and C2, we can say that we have the transonic flow through the outer point and inner point in the region of B1 + C1, and of B2 + C2 respectively, when $0 < \varepsilon_{\rm in} \leq \varepsilon_{\rm in,\; c}$. There exists one saddle-type solution in the region D when $\varepsilon_{\rm in} > \varepsilon_{\rm in,\; c}$ and the distance of the transonic point decreases as the specific angular momentum increases as seen from figure \ref{fig:g=3per2}-(b).}
\label{fig:L2-e}
\end{figure}

From the values of $\lambda_{\rm in}^{2}/\cos^{2} \theta$ at the cross point on the $\lambda_{\rm in}^{2} / \cos^{2} \theta$ - $\Psi$ plane for various $\varepsilon_{\rm in}$ values, we can have the boundary between the respective regions in which the transonic point appears in the inner region and in the outer region, on the $\lambda_{\rm in}^{2}/\cos^{2} \theta$ - $\varepsilon_{\rm in}$ plane, which is shown in figure \ref{fig:L2-e}. 
Figure \ref{fig:L2-e} also reveals the region in which the three-sonic-point solutions exist on the $\lambda_{\rm in}^{2}/\cos^{2} \theta$ - $\varepsilon_{\rm in}$ plane.
The two thick boundary lines are obtained from equations of $d(\lambda_{\rm in}^{2}/\cos^{2} \theta)/dX\; =\; 0$ for various $\varepsilon_{\rm in}$ values and the thick dashed line represents the relation in equation (\ref{eqn:varepsilon_imp}). 

This figure is consistent with the result by Fukue (1987).  It should be noted here that 
Fukue (1987) pointed out a possibility for a shock to appear in the region B1 and to change the path of the supersonic flow through the outer sonic point to a subsonic flow returning to the supersonic flow through the inner sonic point.
This idea was later expanded to the model for the mass ejection behind the shock (e.g. Kumar \& Chattopadhyay 2013; 2017).

Here, let us compare the specific angular momentum, $\ell_{\rm in}$, with that of the Keplerian circular motion, $\ell_{\rm k}$, at the place with a distance, $R$, and an elevation angle, $\theta$, which is given as
\begin{equation}
\ell_{\rm k} = \sqrt{GM} \frac{R^{3/2} \cos^{2} \theta}{R-R_{\rm s}}.
\label{eqn:ell_k}
\end{equation}
If we introduce the ratio parameter, $\omega$, defined as $\omega = \ell_{\rm in}/\ell_{\rm k}$, we can obtain the following equation from equations (\ref{eqn:TrnSncCond_Rin_1}) and (\ref{eqn:ell_k}) as
\begin{equation}
\omega^{2} \cos^{2} \theta = \frac{1}{1-\Gamma} \frac{1}{X} \{ -2 \Gamma \varepsilon_{\rm in} (X-1)^{2} + (1-2\Gamma) X + 2\Gamma\}.
\label{eqn:omega}
\end{equation}
Figure \ref{fig:g=3per2}-(c) shows the distributions of $\omega \cos \theta$ against $X$ for various $\varepsilon_{\rm in}$ values and we see that $\omega \cos \theta$ is close to unity in the inner region, while it is significantly smaller than unity in the outer region.
Since the viscid, subsonic ADAF is considered to tend to have a $\omega$ significantly smaller than unity (Narayan \& Yi 1995), it would be natural that the outer viscid subsonic ADAF turns to the inner inviscid supersonic flow through the outer sonic point corresponding to the lower $\omega$ value, in the three-sonic-point range of $\varepsilon_{\rm in,\; c} \geq \varepsilon_{\rm in} > 0$.

\begin{figure}
\begin{center}
  \includegraphics[width=8cm]{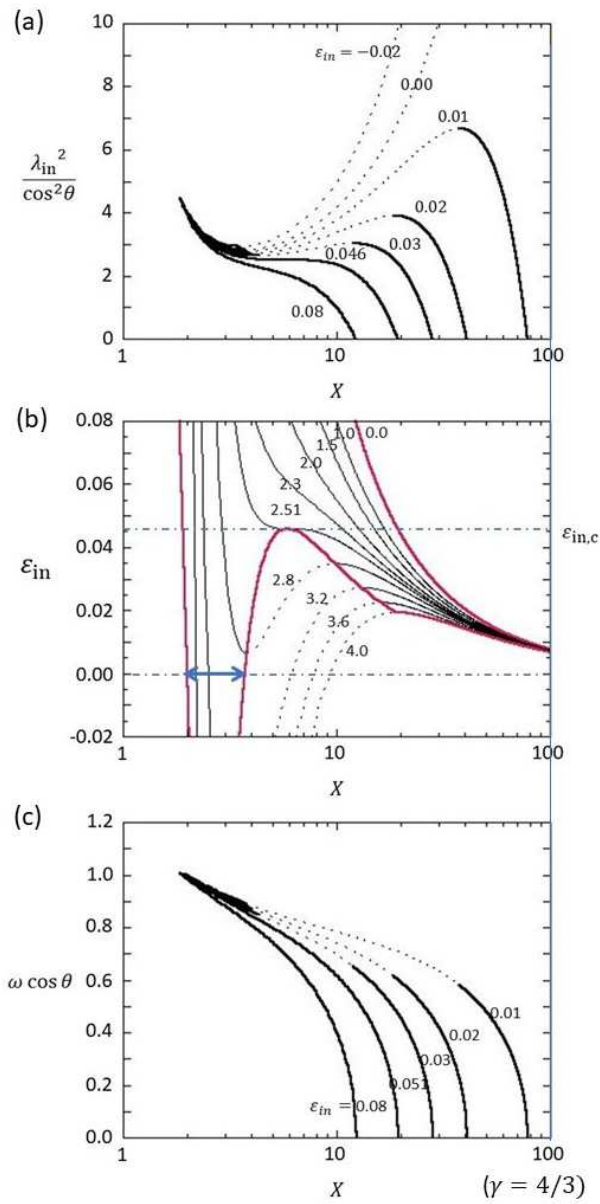} 
\end{center}
\caption{The same diagrams as in figure \ref{fig:g=3per2} but in case of $\gamma = 4/3$.}
\label{fig:g=4per3}
\end{figure}

The same three diagrams as in figure \ref{fig:g=3per2} are presented for $\gamma=4/3$ in figure \ref{fig:g=4per3}.
In this case, $\varepsilon_{\rm in,\; c}$ = 0.046.


\section{Steady jet ejections from the vicinity of black hole}
As shown in section \ref{Introduction}, the steady jet ejections are observed in the ADAF configuration and the slim disk configuration.
In the contexts, we discuss the following two scenarios for steady jet ejections from the vicinity of the black hole.

\subsection{Steady jet ejection in the low/hard state}
We suppose the two-layer situation in the ADAF configurations as schematically drawn in figure \ref{fig:two-layer-configuration} and discuss a scenario for the steady jet ejection in the low/hard state here.

Although the two-layer situation is likely to appear even in the slim disk configuration, the mass flow rate through the upper flow is considered to be significantly smaller than that through the main flow in the slim-disk configuration (Inoue 2021a) and the presence of the upper flow would be negligible in the more powerful jet-ejection process in the main flow of the slim disk, which is discussed later in subsection \ref{JetEjection-SlimDisk}.

\subsubsection{Structure of the main flow}\label{MainFlow}
The main flow is considered to evolve from the standard thin disk on the outer side and thus $\varepsilon_{\rm in}$ should be negative by a minute amount of 10$^{-4}$ or so.
In the case of $\varepsilon_{\rm in} < 0$ as studied in the previous section, the viscid subsonic ADAF inflowing from the outside is considered to turn to the inviscid supersonic flow through the transonic point close to the black hole.
Since the absolute value of $\varepsilon_{\rm in}$ is very small, the range of the transonic position in $X$ can be seen practically from the case of $\varepsilon_{\rm in} = 0$ and is 2 $\sim$ 4.7 for $\gamma = 3/2$ as seen from figure \ref{fig:g=3per2}.
The reduction factor, $\omega$, of the specific angular momentum at the transonic point from that of the Keplerian circular motion is, on the other hand, presented in figure \ref{fig:g=3per2}-(c) and is seen to decrease from unity as $X$ increases from 2.
Here, we set $\cos \theta =1$ for the main flow.
The outer viscid subsonic ADAF is considered to tend to have the lower $\omega$ and thus the transonic point is likely to be selected at the outermost position of the allowed range.
Hence, we can expect that the main flow turns from the viscid subsonic ADAF to the inviscid supersonic flow at $X \sim 5$, and that the supersonic flow gets geometrically thiner and thiner as the Mach number increases in approaching to the black hole (see e.g. Sadowski 2009).

\subsubsection{Structure of the upper flow and jet ejection}\label{JetEjection}
Let us consider, now, what happens to the upper flow above the main flow as it approaches to the black hole.

The matter in the upper flow could tend to merge with the main flow through turbulent motions transferring the angular momentum across the boundary between the two flows.
Then, the viscous transport energy, $\varepsilon_{\rm v}$, outflowing in the main flow could be carried into the upper flow.
If this energy transfer from the main flow to the upper flow really takes place, we can expect a situation for the upper flow to have a transonic point, corresponding to the case of $e_{\rm in} > 0$ as considered in the previous section.

\begin{figure}
\begin{center}
  \includegraphics[width=8cm]{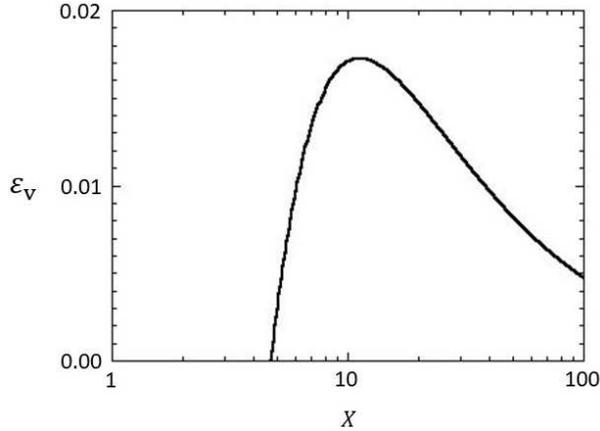} 
\end{center}
\caption{Distribution of $\varepsilon_{\rm v}$ against $X$ calculated from equation (\ref{eqn:e_v-mf}) assuming $\omega$ = 0.8 and $X_{\rm in,\; m}$ = 4.7.}
\label{fig:X-e_v}
\end{figure}

The viscous transport energy, $\varepsilon_{\rm v}$, in the main flow can be expressed as a function of $X$ from equations (\ref{eqn:e_v}) as
\begin{equation}
\varepsilon_{\rm v} = \omega^{2} \frac{X}{(X-1)^{2}} \left\{ 1 - \frac{\omega}{\omega_{\rm in}} \frac{X-1}{X_{\rm in,\; m}-1} \left(\frac{X_{\rm in,\; m}}{X} \right)^{3/2} \right\},
\label{eqn:e_v-mf}
\end{equation}
where $X_{\rm in,\; m}$ represents the position of the transonic point of the main flow and $\omega_{\rm in}$ is $\omega$ at the transonic point.
Figure \ref{fig:X-e_v} shows the $\varepsilon_{\rm v}$ distribution as a function of $X$ calculated assuming $\omega = \omega_{\rm in}=0.8$ and $X_{\rm in,\; m}=4.7$.
If a fraction of this amount of $\varepsilon_{\rm v}$ is transfered to the upper flow, we can roughly expect that the upper flow has a positive specific energy around 0.01.

If the upper flow has $\varepsilon_{\rm in} \sim 0.01$, the viscid subsonic ADAF in the upper flow is expected to turn to the inviscid supersonic flow around $X \sim 20$ as seen from figure \ref{fig:g=3per2}-(a). 

\begin{figure}
\begin{center}
  \includegraphics[width=10cm]{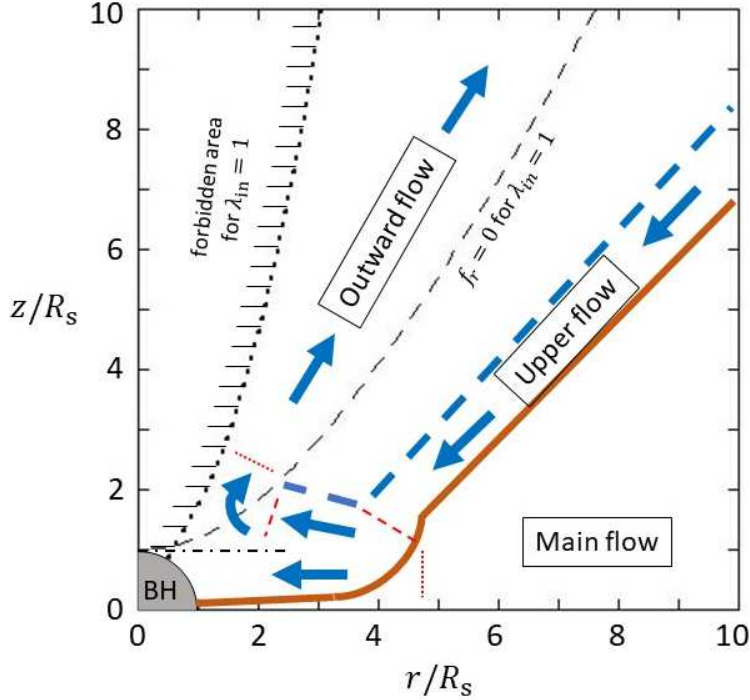} 
\end{center}
\caption{Schematic cross section of the innermost part of the two-layer configuration indicating the jet-ejecting process.  The main flow on the bottom side of the two layers turns to be supersonic at $r/R_{\rm s} \sim 5$ and tends to be geometrically thin as it approaches the black hole.  The supersonic upper flow along the uppermost limb of the main flow forms the oblique shock by colliding to the thin supersonic main flow.  A part of the upper flow behind the oblique shock which enters the region of $z/R_{\rm s} > 1$ is pushed back in the $r$ direction by the centrifugal force and pushed up in the $z$ direction by the thermal pressure enhanced behind the second shock.  As the result, the flow changes to the outward flow and turns to be supersonic at the distance of 2 to 3 $R_{\rm s}$.  Here, the thin dashed line indicates the boundary of $f_{\rm r}=0$ from equation (\ref{eqn:f_r}) for $\lambda_{\rm in}=1$ above which the centrifugal force is dominant to the gravitational force in the $r$ direction, and the hatched region is the forbidden area given by the constraint from equation (\ref{eqn:Cond-lambda}) for $\lambda_{\rm in}=1$.  Thin dotted lines represent the turning positions to the supersonic flow and thin short dashed lines indicate the shock positions.
}
\label{fig:SteadyJetConfiguration}
\end{figure}

The supersonic upper flow advances inward along the uppermost limb of the main flow, while the main flow turns to the supersonic flow at $X \sim 5$ and gets geometrically thin on the closer side to the black hole.
Since the limb of the main flow is considered to change its elevation angle from say 45$^{\circ}$ to almost 0$^{\circ}$ around  $X \sim 3 \sim 4$,  
the kinetic energy of the $v_{\rm z}$ component in the supersonic upper flow should be converted to the thermal energy and the oblique shock is likely to be formed there.
Figure \ref{fig:SteadyJetConfiguration} exhibits the schematic diagram of the situation.

The gas comes to flow in parallel to the equatorial plane on the downstream side of the oblique shock, having a slight expanding motion in the $z$ direction due to the thermalization of the partial kinetic energy.
The force per mass, $f_{\rm r}$, which the gas receives in the $r$ direction in the region behind the oblique shock is given as
\begin{equation}
f_{\rm r} = \frac{GM}{R_{\rm s}^{2}} \left( \frac{2 \lambda_{\rm in}^{2}}{\xi^{3}} - \frac{\xi}{(X-1)^{2} X}\right),
\label{eqn:f_r}
\end{equation}
where $\xi = r/R_{\rm s}$, and the boundary of $f_{\rm r} = 0$ for $\lambda_{\rm in} = 1$ is indicated with the thin dashed line in figure \ref{fig:SteadyJetConfiguration}.

The parameter, $\lambda_{\rm in}^{2}/\cos^{2} \theta$, of the upper flow with $\varepsilon_{\rm in} \sim 0.01$ can roughly be estimated from figure \ref{fig:g=3per2}-(a) or (b), and is seen to distribute around 2 for the outer sonic point of the flow.
If we adopt the typical values of $\lambda_{\rm in}^{2}/\cos^{2} \theta$ and $\theta$ to be 2 and 45$^{\circ}$ respectively, $\lambda_{\rm in}$ is unity.

The centrifugal force becomes dominant to the gravitational force in the region above the boundary of $f_{\rm r} = 0$ in figure \ref{fig:SteadyJetConfiguration} and thus the flow which enters this $f_{\rm r} > 0$ region is braked, forming another shock there.
Then, it is expected that the kinetic energy of the $v_{\rm r}$ component is even converted to the thermal energy behind the shock and the enhanced pressure pushes up the gas in the $z$ direction.

Since the gas has the positive specific energy of $\varepsilon_{\rm in} \sim 0.01$, the gas is likely to start outflowing towards infinity and to get supersonic at some distance.
If we assume that it flows along a hollow cone surface with an elevation angle of $\theta$, the outflow is the reverse case of the accretion flow as just studied in the previous section.
Hence, we can see where the sonic point appears from figure \ref{fig:g=3per2}.

The outflowing gas has the normalized specific angular momentum $\lambda_{\rm in} \sim 1$.
The ejected direction, on the other hand, is constrained by equation (\ref{eqn:Cond-lambda}) and $\theta$ cannot be larger than 60$^{\circ}$.

If $\theta$ is slightly smaller than 60$^{\circ}$, the parameter value of $\lambda_{\rm in}^{2}/\cos^{2} \theta$ becomes slightly smaller than 4.
In this case, the sonic point appears around $X = 2 \sim 3$ as seen from figure \ref{fig:g=3per2} - (a). 
The outflow is considered to get supersonic at such a short distance from the black hole.

The trajectory of the flow tends to be ballistic as the Mach number increases and is expected to finally advance along the line on which the centrifugal force and the gravitational force in the $r$ direction balance with each other.
If we introduce the slant angle of the balance line from the $z$ direction, $\psi=(\pi/2)-\theta$, 
we can obtain the following equation from equation (\ref{eqn:f_r}) as
\begin{equation}
\psi \simeq \left( \frac{2 \lambda^{2}}{X} \right)^{1/4},
\label{eqn:psi}
\end{equation}
when $X \gg 1$.

The possibility for jets to flow between the two boundaries as in figure \ref{fig:SteadyJetConfiguration} was already discussed by several authors, where the boundaries given by $f_{\rm r}$ = 0 and equation (\ref{eqn:Cond-lambda}) were called as the centrifugal barrier and the funnel wall respectively (e.g. Molteni et al. 1996; Chattopadhyay \& Das 2007; Kumar \& Chattopadhyay 2013).

\subsection{Steady jet ejection from the slim disk}\label{JetEjection-SlimDisk}
Another scenario for the steady jet ejection than what is studied above is discussed for the slim disk itself in this subsection.

The optical depth is considered to be sufficiently large for the radiation field to be established in the slim disk.
If the radiation pressure is much stronger than the gas pressure, the mechanical balance in the $z$ direction is approximately written as
\begin{equation}
\frac{dP_{\rm rad}}{dz} \simeq -\rho \frac{GMz}{r^{3}},
\label{eqn:dPrdz}
\end{equation}
where $P_{\rm rad}$ is the radiation pressure.
On the other hand, the photon flux in the $z$ direction, $F_{\rm z}$, is given as
\begin{equation}
F_{\rm z} =-\frac{c}{3\kappa_{\rm T} \rho} \frac{dU_{\rm rad}}{dz},
\label{eqn:PhotonDiffEq}
\end{equation}
where $U_{\rm rad}$ is the radiation energy density and $\kappa_{\rm T}$ is the opacity for the Thomson scattering.
Considering the relation of $P_{\rm rad} = U_{\rm rad}/3$, we obtain from above two equations 
\begin{equation}
F_{\rm z} \simeq \frac{cGM}{\kappa_{\rm T}} \frac{z}{r^{3}}.
\label{eqn:F_z}
\end{equation}
Hence, the radiation flow from the bottom side to the upper side should be induced in this situation.

The cooling time of the matter, $t_{\rm c}$, is determined by the photon diffusion time in the $z$ direction
and is approximately written as
\begin{equation}
t_{\rm c} \simeq \frac{h \tau}{c},
\label{eqn:t_c_Def}
\end{equation}
where $\tau$ is the optical depth over the half-thickness of the disk, $h$, which is given as
\begin{equation}
\tau \simeq \rho h \kappa_{\rm T}.
\label{eqn:tau_Def}
\end{equation}
The slim disk is advection-dominated and thus the cooling time should be longer than the matter flowing time, $t_{\rm f}$, which is approximated as
\begin{equation}
t_{\rm f} \simeq \frac{r}{v_{\rm r}}.
\label{eqn:t_f-Def}
\end{equation}
Hence, it is expected that the radiation flux is transfered from the bottom side to the upper side since the two sides co-moves with each other, but that it stops and is carried inward in the upper layer.

Now, we can draw a picture that the main flow of the slim disk itself has two-layers, the bottom layer and the upper layer, and the energy is transfered from the bottom layer to the upper layer by the radiation flux.

The energy loss rate per mass of the bottom layer, $\dot{e}_{\rm bl}^{(-)}$, could approximately calculated as
\begin{eqnarray}
\dot{e}_{\rm bl}^{(-)} &\simeq& \frac{F_{\rm z}}{\rho h_{\rm bl}} \nonumber \\
&\simeq& \frac{c GM}{\kappa_{\rm T} \rho r^{3}} \nonumber \\
&\simeq& \left(\frac{\dot{M}}{\dot{M}_{\rm E}}\right)^{-1} \left( \frac{h}{r}\right) \left(\frac{v_{\rm r}}{r}\right) \eta c^{2},
\label{eqn:e-dot_bl}
\end{eqnarray}
with the help of equation (\ref{eqn:F_z}) with $z = h_{\rm bl}$ and relations
\begin{equation}
\rho \simeq \frac{\dot{M}}{4\pi r h v_{\rm r}},
\label{eqn:rho-Mdot}
\end{equation}
and
\begin{equation}
\dot{M}_{\rm E} = \frac{L_{\rm E}}{\eta c^{2}},
\label{eqn:Mdot_E-Def}
\end{equation}
where $h_{\rm bl}$ is the half thickness of the bottom layer, $L_{\rm E} = 4\pi cGM/\kappa_{\rm T}$ is the Eddington limit and $\eta$ is the energy conversion rate of the black hole accretion.
Then, the energy gain rate per mass of the upper layer, $\dot{e}_{\rm ul}^{(+)}$, is roughly given as
\begin{equation}
\dot{e}_{\rm ul}^{(+)} \simeq \dot{e}_{\rm bl}^{(-)} \frac{\dot{M}_{\rm bl}}{\dot{M}_{\rm ul}},
\label{eqn:edot_ul-edot_bl}
\end{equation}
where $\dot{M}_{\rm bl}$ and $\dot{M}_{\rm ul}$ are the accretion rate through the bottom layer and upper layer respectively.
The total energy gain per mass of the upper layer, $\Delta e_{\rm ul}$, at a position, $r$, can be calculated as
\begin{eqnarray}
\Delta e_{\rm ul} &\simeq& - \int_{r_{\rm ob,\; s}}^{r} \dot{e}_{\rm ul}^{(+)} \frac{dr}{v_{\rm r}} \nonumber \\ &\simeq& \left(\frac{\dot{M}}{\dot{M}_{\rm E}}\right)^{-1} \left(\frac{\dot{M}_{\rm bl}}{\dot{M}_{\rm ul}}\right)  \left( \frac{h}{r}\right) \ln \left( \frac{r_{\rm ob,\; s}}{r}\right)
\label{eqn:Delta-e_ul}
\end{eqnarray}
with the help of equations (\ref{eqn:edot_ul-edot_bl}) and (\ref{eqn:e-dot_bl}), 
where $h/r$ is assumed to be constant against the position change.
Here, $r_{\rm ob,\; s}$ is the outer boundary of the slim disk.
The non-dimensional amount of $\Delta \varepsilon_{\rm ul} = \Delta e_{\rm ul}/(GM/R_{\rm s}) = 2 \Delta e_{\rm ul} c^{2}$ is
estimated to be $\sim 5 \times 10^{-2}$ for $\eta = 0.1$,  $\dot{M}/\dot{M}_{\rm E} = 10$, $\dot{M}_{\rm bl} = \dot{M}_{\rm ul}$,  $(h/r) = 1$ and $(r_{\rm ob}/r) = 10$.

In this picture, the upper layer can have a positive specific energy of $\varepsilon_{\rm in} \sim 0.05$ and then is likely to change from the viscid subsonic flow to the inviscid supersonic flow at $X \sim 10 \sim 20$ in which the normalized specific angular momentum of $\lambda_{\rm in}^{2}/\cos^{2} \theta \sim 2$ is carried inward, as seen from figure \ref{fig:g=4per3}-(a).
On the other hand, the bottom layer has a negative specific energy and is considered to turn to the supersonic flow at $X \sim 4$ as seen from figure \ref{fig:g=4per3}-(b).
This situation is very similar to the two layer configuration as discussed in the previous subsection and the mass ejection is expected to arise in the similar way as discussed there.

\section{Summary and discussion}
Relativistic jet ejections are often observed from binary X-ray sources and AGNs.
Observations indicate that they are associated with accretion flows on to compact objects and emerge from the innermost regions of the accretion flows.

The specific energy of the jet matter is required to be sufficiently positive at the innermost region as to get to infinity with a relativistic speed, while that of the accreted matter is considered to be slightly negative at the outer boundary of the accretion flow as discussed in section 1.
Hence, there should exist a mechanism for the jet matter to get the specific energy in the course of the accretion flow.
For the mechanism, we have proposed a scenario in which two layers exist in the direction perpendicular to the equatorial plane of the accretion flow and one gets energy from the other so largely as to have sufficient specific energy for the jet ejection.

Two cases have been considered for the two-layer situation.

One is what is expected for the accretion flow from the accretion ring at the outermost end which is discussed by Inoue (2021a).
Inoue (2021a) predicts that a geometrically thick accretion flow and a geometrically thin accretion disk extend from the accretion ring.
The geometrically thin accretion disk is currently considered to turn to the optically thin ADAF on the way to the compact object in the ADAF configuration and to the slim disk in the slim disk configuration.
As the result, such configuration as in figure \ref{fig:two-layer-configuration} is likely to appear.
For the configurations, the possibility for the energy transfer from the main flow to the upper flow through turbulent mixing of the matter between the two layers has been discussed, and the specific energy normalized by $GM/R_{\rm s}$ transfered to the upper flow has been estimated to be around 0.01.
If the matter with this amount of the specific energy is ejected as the jet, the terminal velocity at infinity should get $\sim$ 0.1 $c$.
Since the ADAF configuration corresponds to the low/hard state in the black hole binaries, the jets in this case can explain the weakly relativistic steady jets observed in the low/hard state.

The power of the mass ejections from the upper flow is expected to be proportional to the accretion rate through the upper flow, namely, the optically thin ADAF sandwiching the standard thin disk from the outermost side of the accretion flow.
Inoue (2021a) studied properties of the accretion ring at the outermost edge of the accretion flow and predicted that the ratio of the accretion rate through the thick ADAF to that through the standard disk increases as the total accretion rate decreases.
This predicted relative behavior of the two accretion rates could explain the increase of the jet power relative to that in the X-ray emission in association with the decrease of the mass accretion rate (Fender et al. 2003).


The other case for the two-layer situation has been argued for the slim disk itself, which is the result of the energy transfer by the upward photon diffusion in the disk.
The normalized specific energy transfered is approximately estimated to be $\sim$ 0.05 and the terminal velocity is expected to $\sim$ 0.22 $c$ in this case.
This velocity roughly agrees to the jet velocity observed from SS433.

It should be noted here that no steady jet ejection is observed in the high/soft state.
Several observational evidences indicate that an optically thin ADAF extends close to the central compact object even in the high/soft state (see e.g. section 3 in Inoue 2022a), when no significant evidence of mass ejection is observed (Fender et al. 2004).
This means that simple presence of an ADAF cannot yield the mass ejection.
Something which exists not in the high/soft state but in the low/hard state should play a key role for the steady mass ejection, and it could be the situation of two layer ADAFs.
It is likely to be essential for the jet ejection from the upper flow that the main flow changes from the geometrically thick subsonic ADAF to the geometrically thin supersonic flow near the black hole, which causes the steep drop of the upper flow to the equatorial plane, triggering the jet ejection process.

The process for the jet ejection has been considered commonly to both the two cases of the two-layer situation.
Two steps of conversion of the kinetic energy to the thermal energy are discussed to exist in the jet-ejection process.
The first step is the thermalization of the $v_{\rm z}$ component of the supersonic upper flow through the oblique shock.
This happens by the collision of the supersonic upper flow to the geometrically thin, supersonic main flow.
The remaining $v_{\rm r}$ component is considered to be still supersonic even behind the oblique shock but to be thermalized through another shock for a part of the flow entering the region with $z > R_{\rm s}$ where the centrifugal force becomes dominant to the gravitational force and is likely to completely stop the inward motion.
If all the kinetic energy of the inward motion is converted to the thermal energy there, the thermal energy + the rotational energy is considered to exceed the gravitational energy since the specific energy of the upper flow is positive, and the gas is expected to start outflowing toward infinity.

If the gas outflows along a hollow cone surface with a constant elevation angle, $\theta$, the discussion on the sonic points in the inviscid and adiabatic accretion flow in section \ref{InnermostBoundary} is applicable even to the outflow.
Then, the outflow is inferred to get supersonic soon after the start as discussed in subsection \ref{JetEjection}.
When the Mach number becomes sufficiently large,
the flow is expected to become ballistic and to advance balancing the centrifugal force with the gravitational force in the $r$ direction.
If so, the flow is collimated to have a small opening angle, $\psi$, as given in equation (\ref{eqn:psi}).

If the jet advances close to the Roche lobe, the tidal force from the companion star is likely to weaken the effects of the centrifugal force and the gravitational force within the gravitational territory of the compact object.
If we assume the radius of the territory to be $\sim 10^{6}$ in $X$ and $\lambda_{\rm in}$ of the outflow to be $\sim$ 1, the final opening angle is estimated from equation (\ref{eqn:psi}) to be $\sim 2^{\circ}$, which roughly agrees to that determined from the observation of SS433 (Marshall et al. 2002).


As shown above, the scenarios for the steady jets from the slim disk and from the ADAF in the low/hard state are presented and they are shown to well explain the basic features of the observations.
Since the study is still primitive and leaves many questions to solve, further detailed study is obviously desired.

\begin{ack}
The author appreciates the detailed comments from the referee.
\end{ack}


\end{document}